\documentclass[%
 reprint,
superscriptaddress,
 amsmath,amssymb,
 aps,
pra,
floatfix,
]{revtex4-2}

\usepackage{caption}
\usepackage{subcaption}
\usepackage[version=4]{mhchem}  
\usepackage{graphicx}
\usepackage{xcolor}
\usepackage{graphicx}
\usepackage{dcolumn}
\usepackage{bm}
\usepackage{ulem}

\DeclareRobustCommand{\erase}{\bgroup\markoverwith{\textcolor{red}{\rule[.5ex]{2pt}{0.4pt}}}\ULon}

\usepackage{float}

\definecolor{honey}{HTML}{ec9706}

\usepackage{physics}

\begin{document}

\preprint{APS/123-QED}

\title{Selective Enhancement of Optical Chirality and Spin Angular Momentum in Plasmonic Near-Field}

\author{Naoki Ichiji*}
\affiliation{Institute of Industrial Science, The University of Tokyo, 4-6-1 Komaba, Meguro-Ku, Tokyo 153-8505, Japan}
\email{ichiji@iis.u-tokyo.ac.jp}
\author{Takuya Ishida}
\affiliation{Institute of Industrial Science, The University of Tokyo, 4-6-1 Komaba, Meguro-Ku, Tokyo 153-8505, Japan}
\author{Ikki Morichika}
\affiliation{Institute of Industrial Science, The University of Tokyo, 4-6-1 Komaba, Meguro-Ku, Tokyo 153-8505, Japan}
\author{Daigo Oue}
\affiliation{Metaphotonics Research Team, RIKEN Centre for Advanced Photonics, Saitama 351-0198, Japan}
\affiliation{Instituto de Telecomunica\c{c}\~{o}es, Instituto Superior T\'{e}cnico, University of Lisbon, 1049-001 Lisbon, Portugal}
\affiliation{The Blackett Laboratory, Imperial College London, London SW7 2AZ, United Kingdom}
\author{Tetsu Tatsuma}
\affiliation{Institute of Industrial Science, The University of Tokyo, 4-6-1 Komaba, Meguro-Ku, Tokyo 153-8505, Japan}
\author{Satoshi Ashihara}
\affiliation{Institute of Industrial Science, The University of Tokyo, 4-6-1 Komaba, Meguro-Ku, Tokyo 153-8505, Japan}


\begin{abstract}
The interaction between circularly polarized (CP) light and matter is governed by two fundamental quantities: spin angular momentum (SAM) and optical chirality (OC). While these quantities are inseparable in free space, they can be selectively enhanced in plasmonic near-field regions through appropriately designed structures. We demonstrate that the excitation of circular plasmonic nanostructures with CP light enables selective or simultaneous enhancement of SAM and OC through the excitation of rotating plasmon modes. Electromagnetic field analysis reveals that SAM enhancement originates from transverse SAM induced by unidirectional evanescent waves, whereas OC enhancement is governed by the interference between the plasmonic electric field and incident magnetic field. The finite element method simulations confirm that circular dichroism signals arising from these enhanced near fields clearly depend on the SAM and OC of the local fields, underscoring the importance of structural design in the detection and enhancement of optically active phenomena at the nanoscale.
\end{abstract}
\keywords{plasmonic nanoantenna, plasmonic nanohole, spin angular momentum, optical chirality, circular dichroism}

\maketitle
\section{Introduction}
Circularly polarized (CP) light, characterized by its helical electromagnetic fields, induces unique optical phenomena on interaction with matter. A prominent example is the differential response of a material to left- and right-handed CP light, known as optical activity. This effect encompasses circular dichroism (CD), characterized by a difference in absorption, and circular birefringence (CB), characterized by a difference in refractive index. These phenomena are ubiquitous in natural substances and have been extensively studied, particularly in biomolecular analysis~\cite{Fasman96Book,Hendry10NatNano}, crystallography~\cite{kawasaki05JACS,Niinomi16CrystEngComm}, and magneto–optical (MO) effects~\cite{Faraday1846,Han20AM}.

To describe optical activity phenomena, two key quantities that characterize CP light are commonly employed: spin angular momentum (SAM) and optical chirality (OC). In vacuum, SAM is described by a three-dimensional vector~\cite{Bliokh12PRA,Shi21PNAS}.
\begin{align}
\bm{S} &= \frac{1}{2\omega}\mathrm{Im}(\epsilon_0\bm{E}^{*} \times \bm{E} + \mu_0\bm{H}^{*} \times \bm{H}).\label{Eq:SAM}
\end{align}
The optical chiral density is described by a scalar quantity~\cite{Tang10PRL, Schaeferling12PRX,Mun20LSA}.
\begin{align}
C &= \frac{\epsilon_0}{2} \bm{E} \cdot (\nabla \times \bm{E}) + \frac{\mu_0}{2} \bm{H} \cdot (\nabla \times \bm{H})\nonumber\\
 &= \frac{\omega}{2c^2}\mathrm{Im}(\bm{E}^{*} \cdot \bm{H})\label{Eq:OC}.
\end{align}
Here, $\bm{E}$ and $\bm{H}$ denote electric and magnetic fields, respectively; $\epsilon_0$ and $\mu_0$ denote vacuum permittivity and permeability, respectively; $\omega$ and $c$ denote angular frequency and the speed of light, respectively.

\begin{figure}[b]
  \begin{center}
  \includegraphics[width=8.6cm]{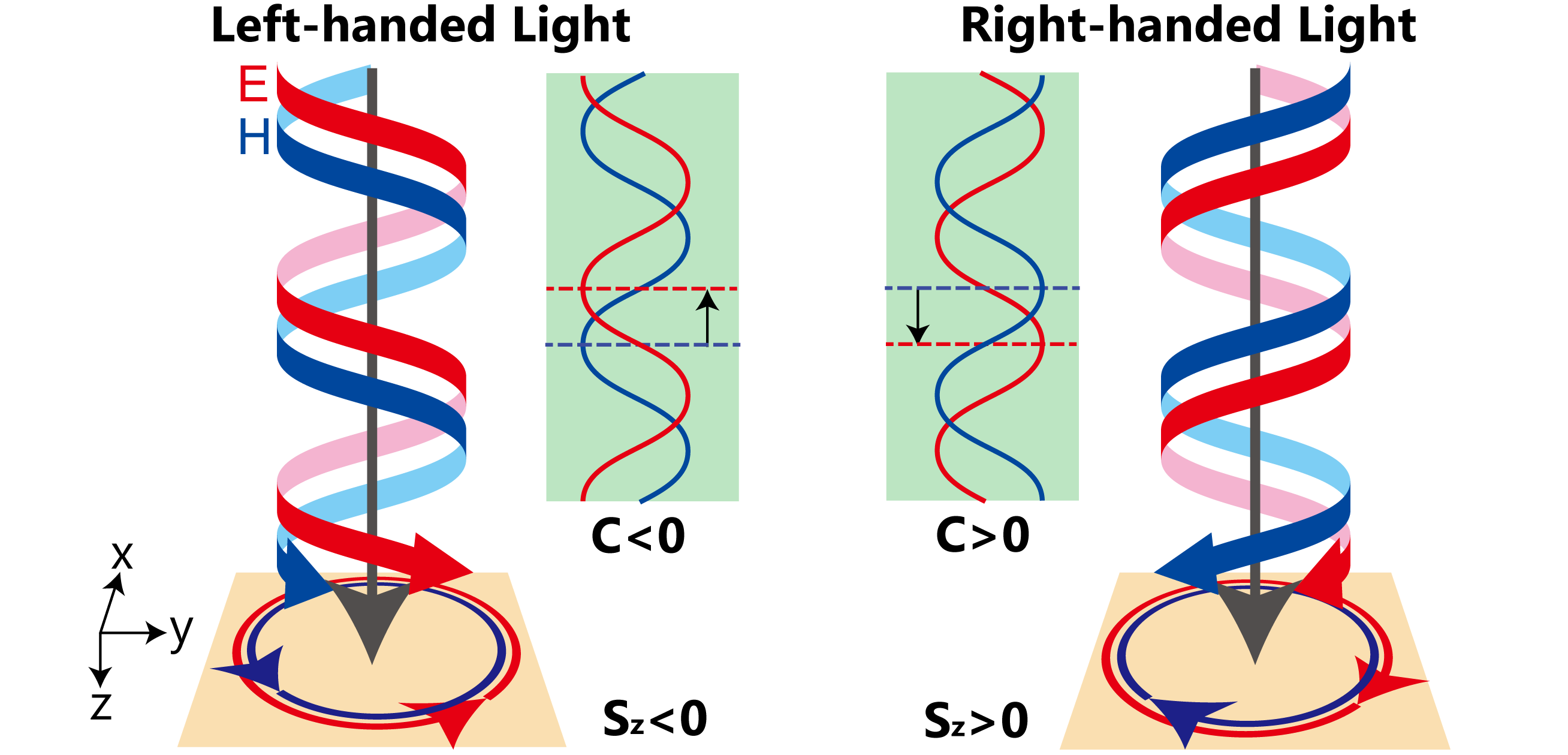}
  \end{center}
  \vspace{-4 mm}
  \caption{Circularly polarized (CP) light propagating in the $z$ direction, exhibiting the electric field $\bm{E}$ (red) and magnetic field $\bm{H}$ (blue) for right- and left-handed polarization. The yellow and green panels represent the characteristic features of SAM and OC, respectively.}
  \label{Fig:image}
\end{figure}

\begin{figure*}[t]
  \begin{center}
  \includegraphics[width=16.8cm]{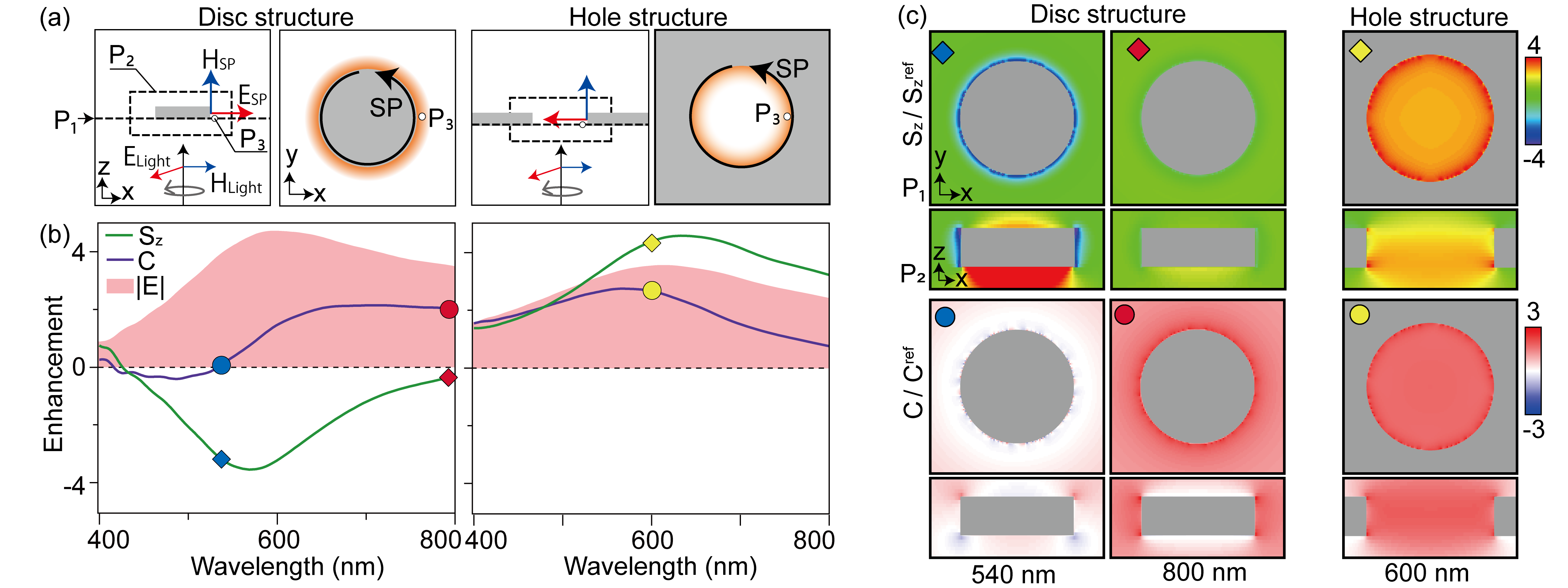}
  \end{center}
  \caption{(a) Finite-difference time-domain (FDTD) simulation model showing the CP light source, plasmonic structures, and the observation planes ($P_{1}$ and $P_{2}$) and point ($P_{3}$). $P_{1}$ and $P_{2}$ are placed in the $xy$-plane at $z = 5$ nm and in the $xz$-plane at $y = 0$, respectively. $P_{3}$ (white circle) is positioned 5 nm away from the wall surface in the same plane as $P_{1}$. The bottom surfaces of the structures are located at $z = 0$, with the radii of the disk and hole structures given by $r_d = 80$ nm and $r_h = 90$ nm. (b) Enhancement spectra of the $z$-component of SAM ($S_{z}(\omega)/S_{z}^{\text{ref}}(\omega)$, green) and OC ($C(\omega)/C^{\text{ref}}(\omega)$, purple) at observation point $P_{3}$. The background represents the normalized electric field enhancement spectrum ($|E|(\omega)/|E^\text{ref}|(\omega)$). (c) Spatial distributions of $S_{z}(\omega)/S_{z}^{\text{ref}}(\omega)$ and $C(\omega)/C^{\text{ref}}(\omega)$ at representative wavelengths, indicated by the color markers in (b).}
  \label{Fig:CPFDTD}
\end{figure*}

Although both are associated with CP light and change sign upon handedness reversal, they describe fundamentally different aspects of the electromagnetic field~\cite{Bliokh11PRA,Coles12PRA,Wu20PRA,Forbes24JOpt}. The vector $\bm{S}$ represents the temporal rotation of the electromagnetic fields, as illustrated in the yellow panels of Fig.~\ref{Fig:image}.
By contrast, $C$ quantifies the phase relationship between the co-directional components of electric and magnetic fields. As shown in the green panels of Fig.~\ref{Fig:image}, the sign of $C$ depends on whether the electric field leads or lags the magnetic field in phase.

As quantities that characterize CP light, SAM and OC have been used to explain various handedness-dependent phenomena in light–matter interactions. However, the physical quantity corresponding to the observed optical activity signal depends on the material properties (Appendix~A). For example, while MO substances and chiral molecules are well-known optically active materials, their responses are governed by SAM and OC, respectively~\cite{Tang10PRL,Mun20LSA,Choi17NC}.
Furthermore, in complex systems, such as surface-adsorbed, oriented chiral molecules, or magnetochiral materials exhibiting nonreciprocal behavior, the optical activity signal exhibits the combined contributions of SAM and OC, rather than being governed by a single quantity~\cite{Tomita18JPD, Chen24NatC}.
In such cases, the individual contributions of SAM and OC to the observed optical activity signals need to be evaluated. However, this task remains challenging, as the two quantities are inherently inseparable in free space~\cite{Coles12PRA,Wu20PRA}.

Herein, we demonstrate the selective enhancement of SAM and OC by exciting unidirectionally rotating plasmon modes through CP light incident on circular metal nanostructures, using near fields enhanced by plasmonic resonance. Specifically, in disk structures, SAM is predominantly enhanced above the resonance frequency, whereas OC is selectively amplified below it. By contrast, hole structures enable the simultaneous enhancement of both quantities at the resonance frequency. By analyzing the three-dimensional electromagnetic field distribution, we elucidate the respective mechanisms responsible for the enhancement of SAM and OC. SAM enhancement is driven by transverse SAM (t-SAM), which arises from the unidirectional rotation of evanescent waves, whereas OC enhancement originates from the interference between the enhanced plasmonic electric field and the incident magnetic field. Finite element method (FEM) simulations incorporating MO and chiral materials confirm the experimental feasibility of enhancing SAM and OC.

\section{FDTD simulation}

Figure~\ref{Fig:CPFDTD}(a) shows the finite-difference time-domain (FDTD) simulation model. The left and right panels depict the disk and hole structures, respectively. A 50-nm-thick Ag structure with radius $r$ is placed on the $xy$-plane, and CP light is normally incident from a total-field scattered-field (TFSF) source. Observation planes ($P_{1}$ and $P_{2}$) and point ($P_{3}$) are defined as shown in the figure. All simulations are performed using ANSYS Lumerical (2023 R1). Details of the simulation conditions are provided in the Appendix~B.

SAM and OC are calculated based on Eqs.~(\ref{Eq:SAM}) and (\ref{Eq:OC}) for each model and normalized to the corresponding values obtained from a reference model without the structure.
Figure~\ref{Fig:CPFDTD}(b) shows the enhancement spectra of the $z$-component of SAM (green, $S_{z}(\omega)/S_{z}^{\text{ref}}(\omega)$) and OC (purple, $C(\omega)/C^{\text{ref}}(\omega)$) at observation point $P_{3}$.
The light red region indicates the normalized electric field enhancement spectrum ($|E|(\omega)/|E^\text{ref}|(\omega)$) at the same point, showing a broad plasmon resonance centered around 600 nm.
The wavelength dependence of $S_{z}$ and $C$ around the disk structure exhibits markedly different behavior: $S_{z}$ displays a sign-inverted profile relative to the electric field enhancement, while $C$ shows a dispersive profile centered near the resonance wavelength.
By contrast, both $S_{z}$ and $C$ for the hole structure exhibit peak shapes similar to that of the electric field enhancement, with broad resonance features centered around 600 nm.

The spatial distributions in the $P_{1}$ plane at typical wavelengths, indicated by the color markers, are shown in Fig.~\ref{Fig:CPFDTD}(c). In the disk structure, at $\lambda \!=\!540$ nm, $S_{z}$ is enhanced along the circular boundary, whereas $C$ is nearly absent despite CP light excitation.
At $\lambda \!=\!800$ nm, the situation is reversed, with $C$ being enhanced and $S_{z}$ nearly vanishing. The hole structure exhibits the simultaneous enhancement of both quantities within the hole region at its resonance wavelength ($\lambda \!=\!600$ nm).
The $xz$-plane cross section at the $P_{2}$ plane (bottom panel) shows that OC retains the same sign throughout the region, unlike typical plasmon-induced OC enhancement, which often exhibits opposite signs across spatial coordinates. Interestingly, this coordinate-dependent sign variation instead appears in $S_{z}$: at 540 nm, $S_{z}$ with opposite signs is clearly visible on the sidewalls and the top and bottom surfaces.

\subsection{Spin angular momentum}
\begin{figure}[t]
  \begin{center}
  \includegraphics[width=8.6cm]{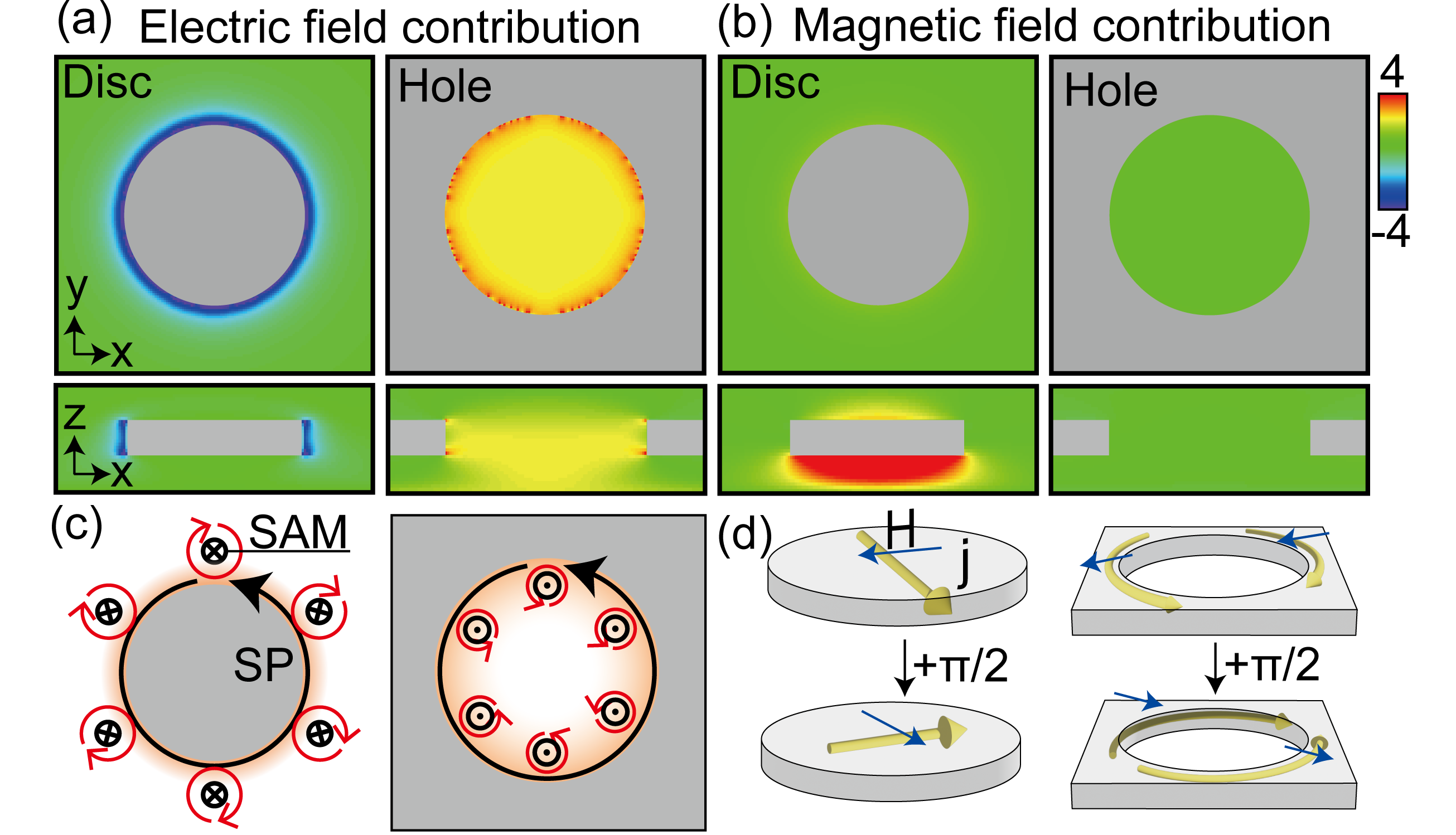}
  \end{center}
  \vspace{-4 mm}
  \caption{Components contributing to $S_{z}$ around plasmonic nanostructures. (a, b) Spatial distributions of the electric and magnetic field contributions to $S_{z}$ for both disk (left) and hole (right) structures: (a) $S^{E}_{z}\!=\!\tfrac{1}{2\omega}\text{Im}(\epsilon_{0}(E^{*}_{x}E_{y}-E^{*}_{y}E_{x}))$ and (b) $S^{H}_{z}\!=\!\tfrac{1}{2\omega}\text{Im}(\mu_{0}(H^{*}_{x}H_{y}-H^{*}_{y}H_{x}))$. (c) Rotational behavior of the electric field associated with the rotating plasmon mode at the wall surface. (d) Schematic diagram of the magnetic field induced by charge motion.}
  \label{Fig:SAM}
\end{figure}

\vspace{-2 mm}
First, we discuss the behavior of SAM. To better understand its complexity, it is useful to separately examine the contributions from the electric and magnetic fields.
Figures~\ref{Fig:SAM}(a, b) present the spatial distributions of the electric and magnetic contributions to $S_{z}$, namely $S^{E}_{z}\!=\!\tfrac{1}{2\omega}\text{Im}(\epsilon_{0}(E^{*}_{x}E_{y}-E^{*}_{y}E_{x}))$ and $S^{H}_{z}\!=\!\tfrac{1}{2\omega}\text{Im}(\mu_{0}(H^{*}_{x}H_{y}-H^{*}_{y}H_{x}))$—for both the disk and hole structures at their resonance wavelength, $\lambda = 600$ nm. These distributions reveal that $S_{z}$ with a reversed sign on the walls of the disk and hole structures is primarily governed by the electric field, whereas the strong $S_{z}$ observed on the planar surface of the disk is predominantly determined by the magnetic field.

The $S_{z}$ sign reversal from the electric field term can be explained by the rotation of the electric field in a unidirectional evanescent wave, known as transverse SAM ($t$-SAM)~\cite{Bliokh12PRA, Bliokh14NC, Aiello15NP, Shi21NP}.
The direction of $t$-SAM in a unidirectional evanescent wave is given by $\bm{S} = \nabla \times \bm{P}$, where the total energy flow density is defined as $\bm{P} = \text{Re}(\bm{E}^{*} \times \bm{H})/2$~\cite{Shi21PNAS,Ichiji23PRA}.
This relationship indicates that, for a given energy flow direction, the SAM direction is determined by the decay direction of the evanescent field. While typical localized surface plasmons consist of counter-propagating plasmon waves whose $t$-SAM cancels out, a unidirectional plasmon mode excited by CP light avoids this cancellation, sustaining a finite and uniform t-SAM localized in the near-field~\cite{Triolo17ACSP}. In the case of the disk and hole structures excited by the same CP light, their evanescent fields decay inward and outward, respectively, while their rotation directions remain identical~\cite{Fu24NL,Xingyu24arXiv, Ichiji24NP}. As a  result, the SAM direction is reversed between the two structures, as illustrated in Fig.~\ref{Fig:SAM}(c).

Conversely, the strong SAM observed on the planar surface of the disk structure due to the magnetic field term can be attributed to the magnetic field induced by the collective oscillatory motion of charges within the metal.
As illustrated in Fig.~\ref{Fig:SAM}(d), the induced magnetic field resulting from charge motion is distributed in the $xy$-plane on the surfaces of both the disk and hole structures.
In the hole structure, the orthogonal components of the in-plane magnetic field are spatially separated and exhibit linear oscillation, whereas in the disk structure, the magnetic field exhibits rotational behavior on the planar surface, directly contributing to SAM.

These results indicate that the electric and magnetic field contributions to SAM may not always be of comparable magnitude in plasmonic near-field, where one may dominate depending on the local field configuration.
Given that the relative contributions of permittivity and permeability to light–matter interactions vary with wavelength and material properties, these results highlight the importance of separately evaluating the electric and magnetic components when analyzing SAM in the near-field.

\subsection{Optical chirality}
\vspace{-4 mm}
To analyze the wavelength dependence of OC, it is important to note that a planar plasmon wave does not exhibit OC even if the plasmonic field possesses SAM because the electric and magnetic fields consistently oscillate in orthogonal planes~\cite{Bliokh12PRA}.
However, in the configuration where a circular structure is excited by normally incident CP light (Fig.~\ref{Fig:CPFDTD}(a)), both the plasmonic electric field ($E_{\text{SP}}$) and the incident magnetic field ($H_{\text{Light}}$) have coaxial components in the $xy$-plane, allowing OC to consider finite values.
In this case, the incident magnetic field cannot serve as a variable in the enhancement of OC as described by Eq.~(\ref{Eq:OC}); rather, the degree of enhancement is determined by two factors related to the plasmonic electric field: its amplitude and phase delay.

\begin{figure}[t]
  \begin{center}
  \includegraphics[width=8.2cm]{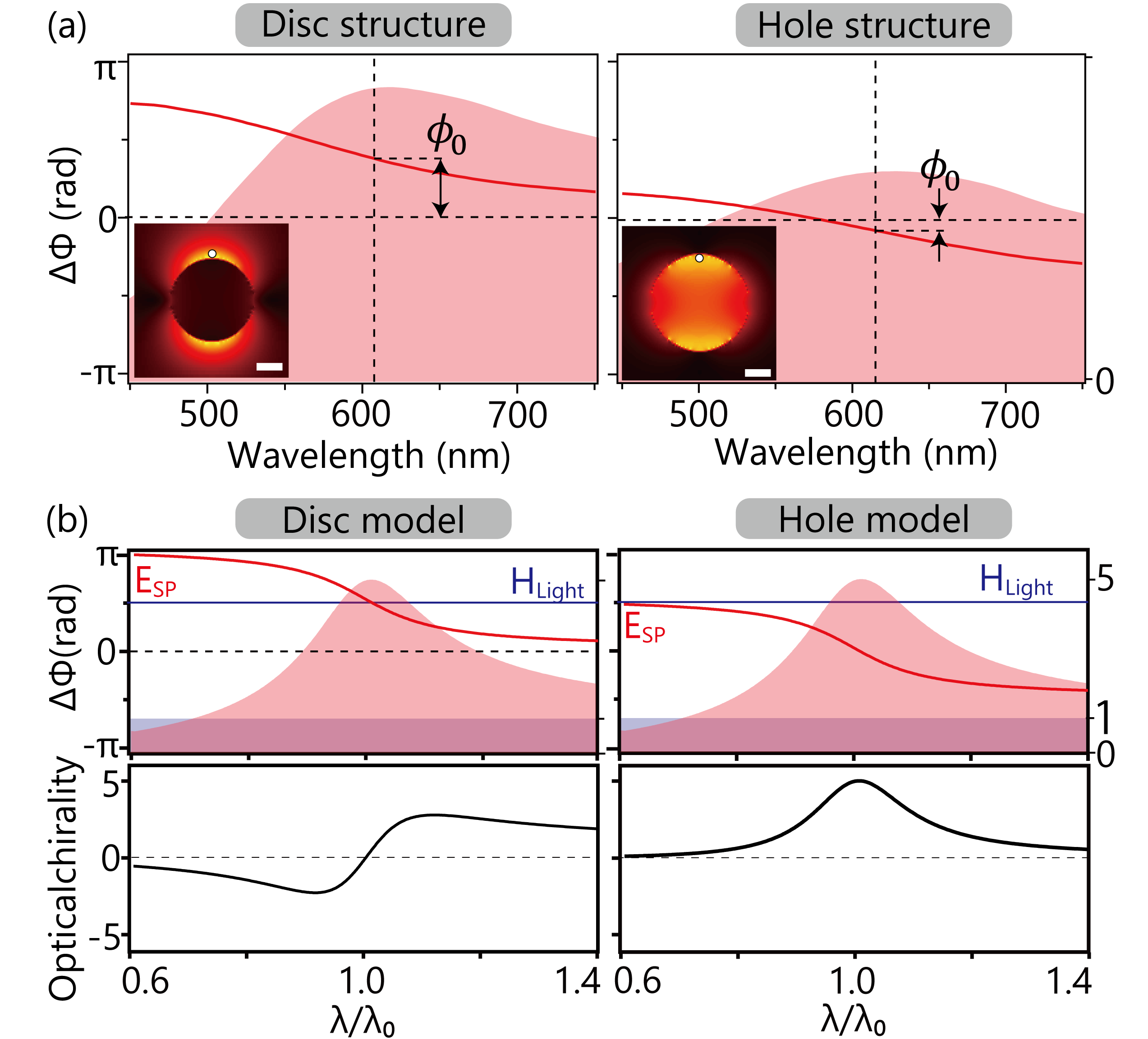}
  \end{center}
  \caption{(a) Phase delay characteristics of the enhanced electric field for the disk (left) and hole (right) structures. The red lines represent the wavelength dependence of the phase delay, and the light red regions indicate the electric field enhancement spectra. The insets show the electric field intensity distributions, $|E|$, at $\lambda\!=\!600$ nm. (b) Simplified model calculation results. (Top panels) Phase difference between the plasmonic electric field $E_{\text{SP}}$ and the incident magnetic field $H_{\text{Light}}$, relative to the incident electric field $E_{\text{Light}}$. The red and blue backgrounds indicate the amplitude spectrum of the electric and magnetic fields, respectively. (Bottom panels) Calculated optical chirality based on the phase-delay characteristics.}
  \label{Fig:delay}
\end{figure}

Figure~\ref{Fig:delay}(a) shows the wavelength dependence of the phase delay characteristics obtained by replacing the CP excitation light with linearly polarized light in the same model as in Fig.~2.
The phase delay is calculated relative to a reference model without a plasmonic structure. Both the disc and hole structures exhibit similar dispersive wavelength dependencies centered at the resonance wavelength, with a noticeable difference only in their baseline offsets.
At resonance, the phase difference $\phi_{0}$ is approximately $\pi/2$ for the disc structure and close to 0 for the hole structure.

This difference in phase-delay characteristics is consistent with previous studies reporting that the resonances of disk-type (positive) and hole-type (negative) structures correspond to electric and magnetic resonances, respectively~\cite{Yang14ACSP, Lee16SciRep, Kihm13OE}.
While the electric resonance is directly driven by the electric field, the charge motion in magnetic resonance is induced by the time variation of the magnetic field. Because the time derivative of a sinusoidal magnetic field leads the field itself by $\pi/2$, this results in a $\pi/2$ phase difference in the charge dynamics between electric and magnetic resonances.
This contrast in phase behavior is also reproduced by an equivalent LCR circuit model, where positive-type and negative-type structures correspond to series and parallel configurations, respectively (Appendix C)~\cite{Engheta05PRL,Zhu14OE,Dubey23PCCP,Schurig09APL, Teperik05PSSA, Withayachumnankul10OE}.

Based on the observed phase relationship between the incident light and plasmonic near-field, we conducted a simplified model calculation comprising the plasmonic electric field ($E_{\text{SP}}$) and the magnetic field of the incident light ($H_{\text{Light}}$). Assuming a classical Lorentz oscillator model for the plasmon resonance, the enhanced spectrum of the amplitude, $A(\omega)$, and the phase delay, $\Delta\phi(\omega)$, are defined by Lorentzian functions as follows~\cite{Goncalves20JPhysD, Oshikiri21ACSNano, Ichiji24PRB}:
\begin{align}
A(\omega) &= \frac{A_{0}}{\sqrt{(\omega_{0}^{2} - \omega^{2})^{2}+(2\gamma \omega)^2}}, \label{Eq:amp}\\
\Delta\phi(\omega) &= \tan^{-1}\left(\frac{2\gamma \omega}{\omega_{0}^{2}- \omega^{2}}\right)+\phi_{0}\label{Eq:phase},
\end{align}
where $A_{0}$, $\gamma$, $\omega$, and $\omega_{0}$ denote the amplitude, damping constant, angular frequency, and resonance angular frequency, respectively. The additional phase $\phi_{0}$ accounts for the differences in the phase delay characteristics between the disk ($\phi_{0} = -\pi/2$) and hole ($\phi_{0} = 0$) structures.
The red lines in Fig.~\ref{Fig:delay}(b) show the phase delay spectrum calculated from Eq.~(\ref{Eq:phase}), using $\omega_{A} = 2\pi \times 500\ \text{THz}$, $\gamma = 2\pi \times 50\ \text{THz}$, and $A_0 = 2.5 \times 10^{5}$. The light red background indicates the amplitude spectrum calculated using Eq.~(\ref{Eq:amp}).

Assuming right-handed CP incident light, the phase difference of $H_{\text{Light}}$ relative to the incident electric field $E_{\text{Light}}$ is set to $\pi/2$ and plotted as a blue line. For simplicity, the incident field intensity is assumed to be frequency-independent, as indicated by the blue background. The OC calculated from these phase relations is shown in the bottom panels, and the resulting spectral trends are consistent with those observed in the FDTD simulations in Fig.~\ref{Fig:CPFDTD}(b).

In the disk model, the $\pi/2$ phase delay compensates for the intrinsic phase difference between the electric and magnetic fields of CP light, causing them to oscillate in phase at the resonance wavelength. As the wavelength deviates from resonance, the lead–lag relationship between $E_{\text{SP}}$ and $H_{\text{Light}}$ reverses, resulting in a sign inversion of OC and producing a dispersive spectral profile centered at the resonance.
By contrast, $E_{\text{SP}}$ in the hole model is enhanced without a phase shift at the resonance frequency, maintaining the $\pi/2$ phase shift between the electric and magnetic fields in CP light.
Although phase shifts occur as the wavelength deviates from resonance, they are not sufficient to reverse the lead–lag relationship between the fields and therefore do not induce a sign change in OC, as observed in the disk model. Consequently, OC exhibits a peak-shaped profile that reflects the electric field enhancement spectrum (Fig.~\ref{Fig:delay}(b)).

Note that this simplified model focuses on $E_{\text{SP}}$ and $H_{\text{Light}}$ in the $xy$-plane; however, in the actual system, additional field components arising from radiation and scattering influence the distribution and intensity of OC~\cite{Cui23ACSP,Biswas24SciAd, Davis13PRB}. However, the phase-delay characteristics between the external and plasmonic fields play a crucial role in explaining the observed differences between the hole and disk structures.

\begin{figure}[t!]
  \begin{center}
  \includegraphics[width=8.6cm]{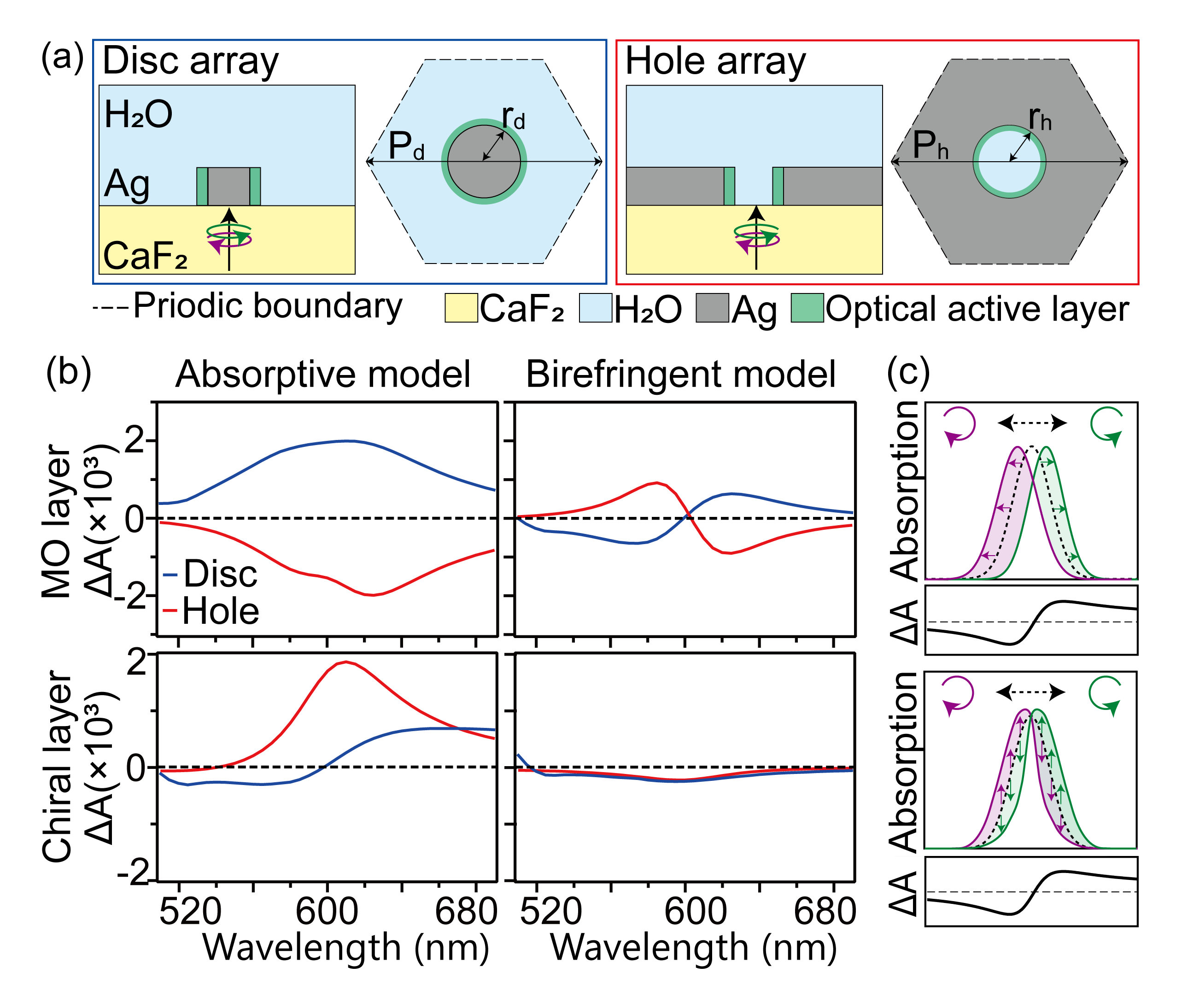}
  \end{center}
  \caption{Schematic and simulation results for FEM simulation. (a) The simulation setup consists of nanodisc and nanohole structures with optically active layers added around their wall surfaces. The disk structure used a radius of $r_d = 105$ nm and a period of $P_d = 380$, while the hole structure used $r_h = 90$ nm and $P_h = 350$. The thickness of all nanostructures was fixed at 50 nm. (b) Circular dichroism (CD) spectra for the disk (blue) and hole (red) structures. The top and bottom panels correspond to magneto–optical (MO) and chiral models, respectively. The left and right columns show the absorptive and birefringent cases. (c) Conceptual illustration of how different CD spectral shapes arise from variations in the absorption spectra.}
  \label{Fig:model}
\end{figure}

\section{FEM simulation}
In the discussion so far, we have shown that SAM and OC in the plasmonic near-field are enhanced based on distinct principles. These findings suggest the potential for independently controlling SAM and OC using plasmonic fields. However, the enhanced electromagnetic fields deviate in several respects from those of free-space CP light, particularly in the amplitude ratio and relative rotation of the electric and magnetic fields. These deviations raise the question of whether the SAM and OC enhanced in this manner can be detected through conventional measurements of optical activity. To evaluate this feasibility and investigate how the enhanced fields manifest in experimentally observable quantities, we conducted FEM simulations incorporating optically active materials.

Figure~\ref{Fig:model}(a) shows the schematic of the FEM simulation setup, conducted using COMSOL Multiphysics. To simulate realistic experimental conditions, the structures were modeled as periodic arrays placed on a substrate in solution. Two types of optically active materials were introduced: a magneto–optical (MO) material responsive to SAM and a chiral material for OC.
The MO material is modeled by modifying the dielectric constant tensor to include off-diagonal components, using an anisotropy coefficient $g$ introduced by magnetization $(\epsilon_{xy} = -\epsilon_{yx} = ig)$~\cite{Li24SA, Kurosawa22OE}. 
For the chiral material, the Pasteur parameter $\kappa$, which governs the coupling between the electric and magnetic responses of the medium, is introduced to incorporate chirality into the system~\cite{Nesterov16ACSP,Mohammadi18ACSP,Mohammadi21ACSP} (Appendix~D).
In bulk materials, the real parts of $g$ and $\kappa$ typically correspond to optical rotation (OR), while the imaginary parts correspond to CD. For $g$, OR manifests as Faraday rotation and CD is referred to as magnetic CD (MCD).
In both cases, the background dielectric constant was set to $\epsilon\!=\!1.43$, and all other simulation conditions were kept identical.

To clarify the fundamental differences in how SAM and OC contribute to optical activity, substantially large constant values were assigned separately to the real and imaginary parts of $g$ and $\kappa$.
In each simulation, only one of these parameters was set to a nonzero value, while the others were kept at zero. The models with nonzero imaginary and real parts of $g$ are referred to as the absorptive ($g = 0.003 i$) and birefringent ($g = 0.003$) MO models, respectively. Similarly, the models with nonzero imaginary and real parts of $\kappa$ are referred to as the absorptive ($\kappa = 0.01 i$) and birefringent ($\kappa = 0.01$) chiral models. Details of the material parameters and simulation settings are provided in the Appendix D.
The light–matter interaction in chiral and MO media is governed by the OC and the electric component of the SAM, respectively (Appendix~A). Accordingly, the coupling parameters $\kappa$ and $g$ are expected to manifest in the CD spectrum—that is, in the absorption difference between right- and left-handed CP light at each frequency.

The CD spectra for each model are shown in Fig.~\ref{Fig:model}(b). Note that although not shown in the figure, reversing the sign of $g$ or $\kappa$ inverts the CD signal, and no CD response is observed when all optical activity parameters are set to zero.
In the absorptive models (left panels of Fig.~\ref{Fig:model}(b)), the CD spectra resemble the wavelength dependence of the near-field $C$ and $S_{z}$ shown in Fig.~\ref{Fig:CPFDTD} (b): the MO models exhibit inverted peak-shaped spectra for the disk and hole structures, whereas the chiral models display a peak-shaped spectrum for the hole structure and a dispersive spectrum for the disk structure. These spectral shapes indicate that the absorption changes at each wavelength directly reflect the enhanced $C$ and $S_{z}$.

The signal for the birefringent chiral model exhibits slight dips for the disk and hole structures. This CD spectrum is consistent with previously reported signals observed in metal nanoparticle–chiral molecule complexes, particularly under off-resonant conditions where the molecular and plasmonic resonances are spectrally separated~\cite{Govorov10NL,Mohammadi23NL,Goerlitzer23ACSP}.
This phenomenon, known as chirality transfer, arises from the coupling between the resonator and surrounding chiral material, causing the system to behave as an effective chiral medium. Therefore, the absorption spectrum does not directly reflect enhanced OC, leading to no notable difference in spectral shape between the disk and hole structures.

By contrast, the CD spectra for the birefringent MO model are clearly inverted between the disk and hole structures, reflecting the modulated effective refractive index due to the oppositely enhanced $S_z$.
Notably, the absorptive chiral model for the disc structure and the birefringent MO models produce similar dispersive CD signals, despite arising from different underlying mechanisms. 
As illustrated in Fig.~\ref{Fig:model}(c), the dispersive CD spectra observed in the MO birefringent models can be attributed to red- or blue-shifts in the resonance wavelengths of the plasmon modes due to modulation of the effective refractive index (top panels), while that in the chiral absorptive model results from a deformation of the absorption spectrum reflecting the enhanced optical chirality (bottom panels).

These results indicate that enhanced OC and SAM generate detectable CD signals, whose spectral characteristics differ between the real and imaginary parts of $g$ and $\kappa$. Simulations with a more practical model—in which both real and imaginary parts coexist—exhibit behavior that appears to be a combination of the individual contributions (Appendix~E), suggesting that the parameter-dependent responses identified in this study provide valuable insights into CD signals in realistic systems.

\section{Conclusion}
In summary, we have demonstrated that the CP excitation of circular plasmonic structures enables the selective control of SAM and OC. Through analysis of their spatial distributions and wavelength dependencies, we clarified that SAM enhancement arises from transverse SAM induced by a unidirectional evanescent field, whereas OC enhancement is governed by the phase relationship between the plasmonic electric field and the incident magnetic field.
FEM simulations incorporating optically active materials confirm that these enhanced near fields can generate CD signals, with spectral characteristics determined by both the plasmonic structure and the nature of the optically active material. 

Our findings suggest that plasmonic near fields would serve as effective platforms for disentangling optically active signals in complex systems, enabling the identification of SAM- and OC-derived contributions and the separate estimation of the real and imaginary parts of $g$ and $\kappa$.
Furthermore, such platforms could allow for detailed analysis of the underlying mechanisms, such as the respective roles of electric and magnetic SAM, and the influence of the electric-to-magnetic field ratio on OC in CD signals.
These capabilities provide a solid foundation for interpreting complex optical activity phenomena.

\section*{Funding Sources}
This work was supported by a Grant-in-Aid for JSPS Fellows (No. JP23KJ0355), a Grant-in-Aid for Scientific Research (A) (No. JP20H00325), a Grant-in-Aid for Scientific Research (C) (No. JP24K08359), a Grant-in-Aid for Challenging Exploratory Research (No. JP20K20560), Fujikura Co., Ltd., and JST, PRESTO (No. JPMJPR24L3). D.O. was supported by a JSPS Overseas Research Fellowship, the Institution of Engineering and Technology (IET), and by Fundação para a Ciência e a Tecnologia and Instituto de Telecomunicações under project UIDB/50008/2020. Moreover, D.O. was supported by the RIKEN special postdoctoral researcher program. 

\section*{Acknowledgments}
The authors thanked A. Kubo and K. Kihara for valuable discussions on the contributions of the electric and magnetic field terms to SAM.

\appendix

\renewcommand\thefigure{\thesection\arabic{figure}} 
\renewcommand\theequation{\thesection\arabic{equation}} 
\setcounter{figure}{0}  
\setcounter{equation}{0}

\section{Theoretical background}\label{Sec: theory}
In magneto–optical (MO) materials, electric polarization acquires an additional component owing to magnetization, often expressed as $\bm{P_e} \propto \bm{M} \times \bm{E}$.
This leads to an interaction energy of the form $\bm{E}^* \cdot (\bm{M} \times \bm{E}) \propto (\bm{E}^* \times \bm{E}) \cdot \bm{M}$, which directly couples the electric SAM to the magnetization $\bm{M}$. This interaction originates from the off-diagonal elements in the permittivity tensor and reflects anisotropy induced by $\bm{M}$.

To describe the interaction between chiral media and electromagnetic fields, we consider the effective dipole moments induced in the material. The electric and magnetic dipole responses are coupled and can be expressed as
\begin{align}
\begin{pmatrix}
\bm{p} \\
\bm{m}
\end{pmatrix}
=
\begin{pmatrix}
\alpha_e & -i\xi \\
i\xi & \alpha_m
\end{pmatrix}
\begin{pmatrix}
\bm{E} \\
\bm{H}
\end{pmatrix},
\end{align}
where $\xi$ represents the chirality of the medium.
The resulting interaction energy is given by
\begin{align} 
&\begin{pmatrix}
\bm{E} \\
\bm{H}
\end{pmatrix}^*
\cdot
\begin{pmatrix}
\bm{p} \\
\bm{m}
\end{pmatrix}
=
\alpha_e \abs{\bm{E}}^2
+ \alpha_m \abs{\bm{H}}^2
+ 2\xi \Im\qty(\bm{E}^* \cdot \bm{H}).
\end{align}
The last term represents a chirality-dependent light–matter interaction and is proportional to the optical chirality $C \propto \mathrm{Im}(\bm{E}^* \cdot \bm{H})$.

\section{FDTD Simulation}\label{Sec: FDTD}
The electromagnetic field distributions were obtained using finite-difference time-domain (FDTD) simulations performed with ANSYS Lumerical FDTD (version 2023 R1).
The simulation model comprised either a 50-nm-thick silver (Ag) disk or hole structure placed on the $xy$-plane, with the radius $r$ set identically for both structures. The surrounding medium was vacuum, and CP light was normally incident along the $z$ direction using a total-field scattered-field (TFSF) source located 5 $\mu$m above the structure.
The computational domain was enclosed by perfectly matched layer (PML) boundary conditions in all directions, and the minimum mesh size was set to 4 nm. The dielectric function of Ag was taken from the Palik optical constants database~\cite{Palik}.
To extract spatial field distributions, two observation planes were defined: $P_{1}$ (in the $xz$-plane at $y = 0$) and $P_{2}$ (in the $xy$-plane at $z = 5$ nm), along with an observation point $P_{3}$ located 5 nm away from the sidewall of the nanostructure.

\section{Circuit-Based Interpretation of Phase Delay}\label{Sec: LCR}
The phase-delay characteristics observed between the disk and hole structures, discussed in the main text in terms of electric and magnetic resonances, can also be interpreted using an equivalent LCR circuit model, a common approach for describing plasmonic resonators.

Considering this analogy, the localized electric field concentrated at the edge of the disk or aperture corresponds to the capacitive element. Regions with a positive real part of the dielectric constant contribute to capacitive behavior, whereas those with a negative real part behave inductively~\cite{Engheta05PRL}. The resistive component accounts for ohmic losses in the metal. Positive-type structures, where charge motion is confined to a single loop through inductive and capacitive regions, are analogous to series LCR circuits~\cite{Zhu14OE,Lu19SR,Dubey23PCCP}. By contrast, negative-type structures, where charge can traverse multiple paths through the slit capacitance and the surrounding inductive metal, correspond to parallel LCR circuits~\cite{Schurig09APL,Teperik05PSSA,Withayachumnankul10OE}.

In the analogy between the plasmonic structure and the LCR circuit model, the localized electric field concentrated between the metal regions at the disk edge or aperture corresponds to the capacitive element. To examine the phase relationship between the plasmonic field and external excitation, we consider the phase difference between the voltage across the capacitor and externally applied voltage. We define the phase delay $\Delta\phi$ as the argument of their complex voltage ratio.

\begin{align}
\Delta\phi(\omega) = \arg\left[ \frac{V_C(\omega)}{V_\text{in}(\omega)} \right],
\end{align}
where $\arg[z]$ denotes the phase angle of a complex number $z$ within the range $(-\pi, \pi)$.

\vspace{3 mm}
\noindent
\textbf{Disk Model (Series LCR): }
In the disk model, the system is represented as a series LCR circuit, and the total impedance observed by the capacitive branch is given by
\begin{align}
Z_{\perp}(\omega) = i\left( \omega L - \frac{1}{\omega C} \right) = \frac{i(\omega^2 LC - 1)}{\omega C}.
\end{align}
The phase difference between the capacitor voltage and the input voltage is expressed as follows:
\begin{align}
\Delta \phi_{\perp}(\omega) = \arg\left[ \frac{Z_C}{R + Z_{\perp}(\omega)} \right],
\end{align}
where $Z_C = -i/(\omega C)$.

Due to the frequency-dependent nature of $Z_{\perp}(\omega)$, we can evaluate $\Delta\phi_{\perp}$ in the three limiting regimes as follows:

\begin{table}[h]
\centering
\caption{Frequency dependence of phase delay $\Delta\phi_{\perp}$ for the series LCR (disk) model.}
\begin{tabular}{l|c|c|c|c}
\hline
Regime & $Z_{\perp}(\omega)$ & $\arg[Z_C]$ & $\arg[R + Z_{\perp}]$ & $\Delta\phi_{\perp}$ \\
\hline
$\omega \ll \omega_0$ & $-i/(\omega C) \rightarrow -\infty$ & $-\pi/2$ & $-\pi/2$ & $0$ \\
$\omega = \omega_0$ & $0$ & $-\pi/2$ & $0$ & $\pi/2$ \\
$\omega \gg \omega_0$ & $i\omega L \rightarrow \infty$ & $-\pi/2$ & $\pi/2$ & $-\pi$ \\
\hline
\end{tabular}
\end{table}

In each regime summarized in the table, $|Z_{\perp}| \gg R$ holds, allowing $R$ to be neglected in $\arg[R + Z_{\perp}]$. Since the capacitor term $Z_C = -i/(\omega C)$ always has phase $-\pi/2$, the phase delay is determined by the phase of the denominator. The resulting values are $0$ at low frequencies, $\pi/2$ at resonance, and $-\pi$ at high frequencies.

In each regime summarized in the table, $|Z_{\perp}| \gg R$ holds, allowing $R$ to be neglected in $\arg[R + Z_{\perp}]$. Since the capacitor term $Z_C = -i/(\omega C)$ always has a phase value of $-\pi/2$, the phase delay is determined by the phase of the denominator. The resulting values are zero at low frequencies, $\pi/2$ at resonance, and $-\pi$ at high frequencies.

\vspace{3 mm}
\noindent
\textbf{Hole Model (Parallel LCR): }
In the hole model, the system is represented as a parallel LCR circuit, and the total impedance is given by
\begin{align} Z_{\parallel}(\omega) = \frac{1}{i\omega C - \frac{1}{i\omega L}} = \frac{i\omega L}{1 - \omega^2 LC}. 
\end{align} 
The phase difference between the capacitor voltage and the input voltage is expressed as follows:
\begin{align} 
\Delta \phi_{\parallel}(\omega) = \arg\left[ \frac{Z_{\parallel}(\omega)}{R + Z_{\parallel}(\omega)} \right]. 
\end{align}
The frequency dependence of $Z_{\parallel}(\omega)$ differs fundamentally from the series case: away from resonance, $Z_{\parallel}$ becomes small in magnitude and the resistance $R$ dominates in the denominator. The resulting phase behavior in each regime is summarized in the table below.

\begin{table}[h]
\centering
\caption{Frequency dependence of phase delay $\Delta\phi_{\parallel}$ for the parallel LCR (hole) model.}
\begin{tabular}{l|c|c|c|c}
\hline
Regime & $Z_{\parallel}(\omega)$ & $\arg[Z_{\parallel}]$ & $\arg[R + Z_{\parallel}]$ & $\Delta\phi_{\parallel}$ \\
\hline 
$\omega \ll \omega_0$ & $i\omega L \rightarrow +0$ & $\pi/2$ & $0$ & $\pi/2$ \\
$\omega = \omega_0$ & $\infty$ & $\pi/2$ & $\pi/2$ & $0$ \\
$\omega \gg \omega_0$ & $-i/(\omega C) \rightarrow -0$ & $-\pi/2$ & $0$ & $-\pi/2$ \\
\hline
\end{tabular}
\end{table}

As shown in the table, in the off-resonant regimes ($\omega \ll \omega_0$ and $\omega \gg \omega_0$), $|Z_{\parallel}| \ll R$ holds, making the resistance dominant in the denominator and keeping $\arg[R + Z_{\parallel}] \approx 0$. At resonance, the impedance diverges, and the phase of both numerator and denominator approaches $\pi/2$, resulting in a net delay of zero. The total phase variation is limited to $\pm\pi/2$.

\section{FEM Simulation}\label{Sec: FEM}
The finite element method (FEM) simulations were performed using COMSOL Multiphysics. The model comprised a silver (Ag) nanodisc or nanohole embedded in a homogeneous, nondispersive medium with a dielectric constant of $\epsilon = 1.43$. CP light was incident normally from the bottom along the $z$ direction, and periodic boundary conditions were applied in the $xy$-plane to simulate an infinite array. To align the resonance wavelengths of the two geometries, the radius and periodicity were designed for each structure: the disk structure had a radius of $r_d = 105$ nm and a period of $P_d = 380$ nm, while the hole structure exhibited $r_h = 90$ nm and $P_h = 350$ nm. The thickness of all nanostructures was fixed at 50 nm.

To examine optical activity, two types of optically active media were introduced: a MO material and a chiral material. Both were modeled, as thin layers conformally coating the sidewalls of the nanostructures. Electromagnetic fields were calculated in the frequency domain, and the CD spectra were obtained by evaluating the difference in absorptance under left- and right-handed CP illumination.

\vspace{3 mm}
\noindent
\textbf{MO effect:}
The MO response arises from the difference in the permittivity for left- and right-handed circular polarizations. Let us define the permittivities for the left- and right-handed polarizations, $\epsilon _ \pm = \epsilon \mp g/2$, where $g$ represents the deviation of the permittivity owing to the MO effect. Here, we assumed that the right-handed (left-handed) polarization experiences lower (higher) permittivity. The permittivity tensor can be explicitly defined as follows~\cite{Li24SA, Kurosawa22OE}.:

\begin{align}
&\boldsymbol{\epsilon}_{\mathrm{MO}}
=
\epsilon _ + \bm{\varepsilon} _ + \bm{\varepsilon} _ + ^ \dagger
+ \epsilon _ - \bm{\varepsilon} _ - \bm{\varepsilon} _ - ^ \dagger
+ \epsilon \bm{u} _ z \bm{u} _ z ^ \dagger
=\begin{pmatrix}
\epsilon & ig & 0 \\
-ig & \epsilon & 0 \\
0 & 0 & \epsilon
\end{pmatrix},
\end{align}
where we have defined the polarization vectors $\bm{\varepsilon} _ \pm = (\bm{u} _ x \pm i\bm{u} _ y)/\sqrt{2}$. The real and imaginary parts of $g$ represent the magnetic CB and magnetic CD, respectively. In particular, we considered $g = 0.003i$ and $g = 0.003$ in the simulation with $\epsilon = 1.43$.

\vspace{3 mm}
\noindent
\textbf{Chirality:}
To incorporate a chiral medium into the simulations, we modeled the material using the Pasteur parameter $\kappa$, which characterizes the coupling between the electric and magnetic responses of the medium. This was achieved by modifying the standard COMSOL optics simulation environment to implement the chiral constitutive relations~\cite{Nesterov16ACSP,Mohammadi18ACSP,Mohammadi21ACSP}:

\begin{align}
\bm{D} &= \epsilon \bm{E} - i\kappa \sqrt{\epsilon_0 \mu_0} \bm{H}, \\
\quad
\bm{B} &= \mu \bm{H} + i\kappa \sqrt{\epsilon_0 \mu_0} \bm{E},
\end{align}
where $\epsilon_0$ and $\mu_0$ are the permittivity and permeability of vacuum, respectively.
The dielectric constant was fixed at $\epsilon = 1.43$, and the permeability was set to $\mu = 1$. The absorptive and birefringent chiral cases were simulated using $\kappa = 0.01i$ and $\kappa = 0.01$, respectively.

To isolate the individual contributions from each optical activity mechanism, $g$ and $\kappa$ were assumed to be either real or imaginary constants, with no frequency dispersion. In each simulation, only one parameter was assigned a finite value, while all others were set to zero.

\section{FEM Simulations with Complex Parameters}
\begin{figure}[h]
  \begin{center}
  \includegraphics[width=8.6cm]{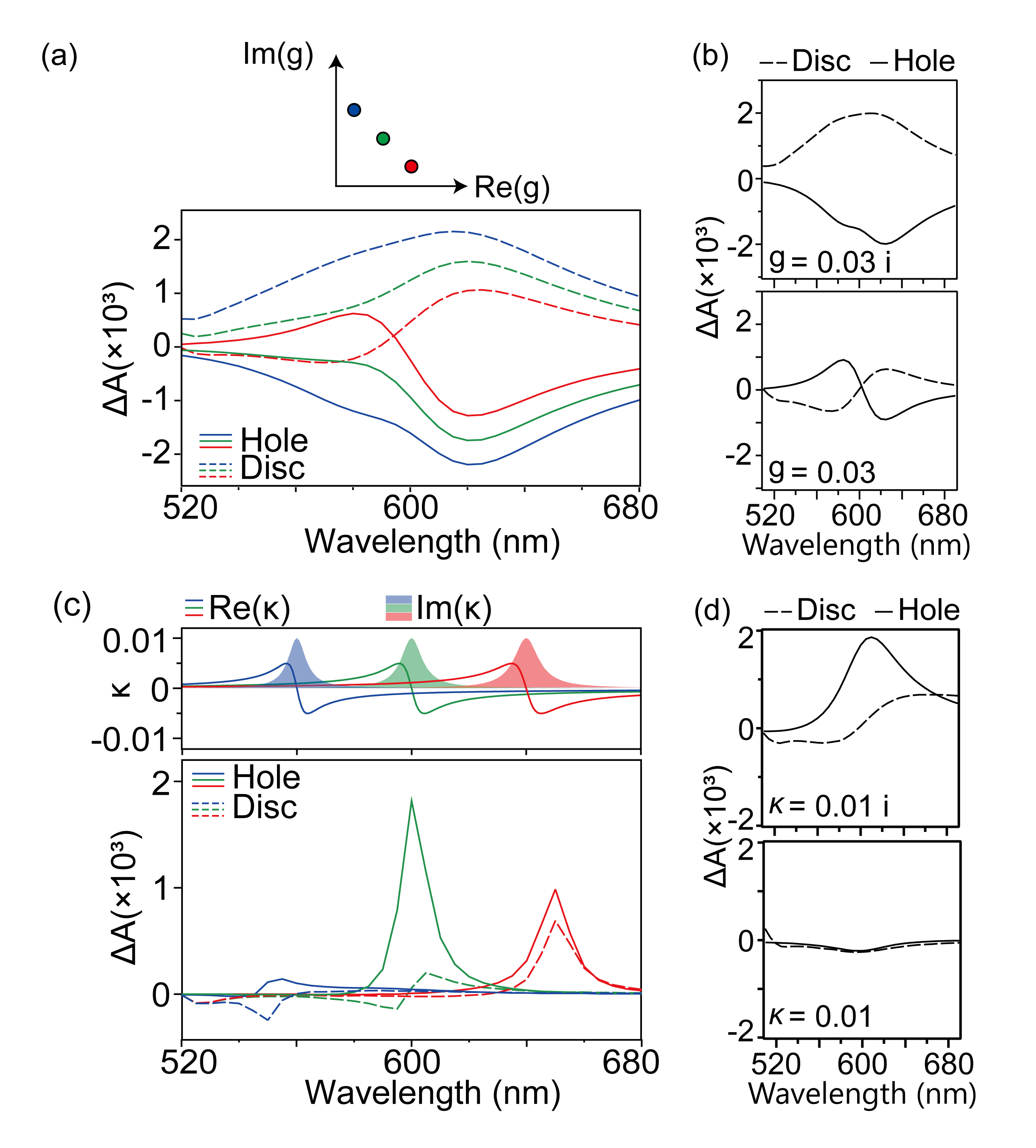}
  \end{center}
  \caption{FEM simulation results for optically active systems with complex-valued $g$ and $\kappa$. (a) CD spectra of model magneto-optical (MO) materials with complex $g = g' + ig''$ values set to $0.001 + 0.005i$ (blue), $0.003 + 0.003i$ (green), and $0.005 + 0.001i$ (red). Solid and dashed lines represent the CD spectra for hole and disc structures, respectively. (b) CD spectra for the idealized constant-$g$ models used in the main text (Fig. 5), shown here for comparison. The upper and lower panels correspond to the absorptive ($g = 0.003i$) and birefringent ($g = 0.003$) cases, respectively. (c) CD spectra for chiral molecules modeled by a Lorentz-type wavelength-dispersive $\kappa(\lambda)$ (Eq. E1), with resonance wavelengths $\lambda_0 = 560$ nm (blue), 600 nm (green), and 640 nm (red). The upper panel shows the real part (solid lines) and imaginary part (shaded regions) of $\kappa(\lambda)$; the lower panel shows the corresponding CD spectra for hole (solid) and disc (dashed) structures. (d) CD spectra for the constant-$\kappa$ models used in the main text (Fig. 5), shown here for comparison. The upper and lower panels correspond to the absorptive ($\kappa = 0.01 i$) and birefringent ($\kappa = 0.01$) models, respectively.}
  \label{Fig:placticalmodel}
\end{figure}

In the FEM simulations presented in this paper, we employed a highly idealized model in which either the real or the imaginary component of $g$ and $\kappa$ is constant non-zero value, while the others are set to zero. This approach was adopted to enable a clear evaluation of the individual contributions of each parameter to the CD spectrum.  
In plactical systems, however, the real and imaginary parts of each parameter generally coexist, and the non-zero component is typically accompanied by a non-negligible counterpart.  
To investigate the CD response under these more complex conditions, we performed simulations based on a complex model that includes both real and imaginary parts of $g$ and $\kappa$.

Figure~E1(a) shows the CD spectra of model MO materials calculated with complex values of $g = g' + i g''$ set to $0.001 + 0.005i$ (blue), $0.003 + 0.003i$ (green), and $0.005 + 0.001i$ (red); all other parameters were kept constant. The lower panel displays the resulting CD spectra for the hole (solid lines) and disc (dashed lines) structures, which exhibit distinct spectral responses depending on the ratio of the real and imaginary parts of $g$. 
For comparison, the CD spectra obtained using the constant-$g$ models shown in Fig.~5 of the main text are replotted in Fig.~S1(b). The upper and lower panels correspond to the absorptive ($g = 0.003i$) and birefringent ($g = 0.003$) MO models, respectively.

These results indicate that the CD spectra for mixed-composition MO models can be interpreted as a combination of the responses induced by the real and imaginary components individually. This observation is consistent with the separate contributions shown in Fig.~4(b) of the main text, and highlights that the complex behaviour in complex systems may be understood as a superposition of idealised limiting cases.

Figure~E1(c) presents simulations that incorporate a molecular layer whose Pasteur parameter $\kappa(\lambda)$ is described by a single-resonance Lorentz model~\cite{Mohammadi18ACSP}:
\begin{equation}
  \kappa(\lambda)=
    \beta\left(
      \frac{1}{\hbar\bigl(c_0/\lambda-\omega_0\bigr)-i\gamma}
      -
      \frac{1}{\hbar\bigl(c_0/\lambda+\omega_0\bigr)-i\gamma}
    \right),
\end{equation}
where $c_0$ is the speed of light, $\hbar$ is the reduced Planck constant, and $\omega_0 = 2\pi c_0 / \lambda_0$ is the molecular resonance angular frequency.  
The scalar coefficient $\beta$ is set to $3.0 \times 10^{-5}$, and the damping constant $\gamma$ is 0.003\,eV.

The upper panel of Fig.~S1(c) shows the wavelength-dependent real part (solid lines) and imaginary part (shaded areas) of $\kappa(\lambda)$, calculated for three models with resonance wavelengths $\lambda_0 = 560$\,nm (blue), 600\,nm (green), and 640\,nm (red).
The corresponding CD spectra are shown in the lower panel, where the hole and disc structures are plotted as solid and dashed lines, respectively.
For comparison, the CD spectra obtained using constant-$\kappa$ models (shown in Fig.~5 of the main text) are replotted in Fig.~S1(d).

Under the present simulation conditions, the CD spectra of the chiral models are predominantly governed by the imaginary part of $\kappa$, leading to more pronounced differences between the disc and hole geometries.
This behaviour can be attributed to the interplay between the spectral shape of $\operatorname{Im}\kappa(\lambda)$ and the enhancement profile of optical chirality (OC).
In the hole structure, where the OC enhancement spectrum exhibits a resonant peak, the resulting CD spectrum shows a sharp peak that reflects the overlap of the two contributions.
In contrast, in the disc structure, where the OC enhancement spectrum is dispersive, the CD spectrum exhibits a negative peak on the short-wavelength side, a positive peak on the long-wavelength side, and a dispersive zero-crossing at the molecular resonance.

\newpage
\bibliography{apssamp}

\end{document}